\documentclass[
 aip,
 amsmath,amssymb,
preprint
]{./revtex4-2}
\usepackage{hyperref}
\usepackage{natbib}
\usepackage{graphicx}
\usepackage{dcolumn}
\usepackage{bm}
\usepackage{multirow}
\usepackage{amsmath}
\usepackage{makecell}
\usepackage{mathrsfs}
\usepackage{tikz}

\begin{document}

\preprint{ArXiv/manuscript}
\pagestyle{plain}

\title{Flow Reconstruction Using Spatially Restricted Domains Based on Enhanced Super-Resolution Generative Adversarial Networks}

\author{Mustafa Z. Yousif}
\affiliation{School of Mechanical Engineering, Pusan National University, 2, Busandaehak-ro 63beon-gil, Geumjeong-gu, Busan, 46241, Rep. of KOREA}

\author{Dan Zhou}
\affiliation{School of Mechanical Engineering, Pusan National University, 2, Busandaehak-ro 63beon-gil, Geumjeong-gu, Busan, 46241, Rep. of KOREA}

\author{Linqi Yu}
\affiliation{School of Mechanical Engineering, Pusan National University, 2, Busandaehak-ro 63beon-gil, Geumjeong-gu, Busan, 46241, Rep. of KOREA}

\author{Meng Zhang}
\affiliation{School of Mechanical Engineering, Pusan National University, 2, Busandaehak-ro 63beon-gil, Geumjeong-gu, Busan, 46241, Rep. of KOREA}

\author{Arash Mohammadikarachi}
\affiliation{School of Mechanical Engineering, Pusan National University, 2, Busandaehak-ro 63beon-gil, Geumjeong-gu, Busan, 46241, Rep. of KOREA}

\author{Jung Sub Lee}
\affiliation{Biomedical Research Institute, Department of Orthopaedic Surgery, Pusan National University Hospital, 179 Gudeok-Ro, Seo-gu, Busan, 46241, Rep. of KOREA}

\author{Hee-Chang Lim}
\email[]{Corresponding author, hclim@pusan.ac.kr}
\thanks{}
\affiliation{School of Mechanical Engineering, Pusan National University, 2, Busandaehak-ro 63beon-gil, Geumjeong-gu, Busan, 46241, Rep. of KOREA}
\email{hclim@pusan.ac.kr}

\date{\today}

\begin{abstract}
\small
\vspace{-2em}

This study aims to reconstruct the complete flow field from spatially restricted domain data by utilizing an Enhanced Super-Resolution Generative Adversarial Network (ESRGAN) model. The difficulty in flow field reconstruction lies in accurately capturing and reconstructing large amounts of data under nonlinear, multi-scale, and complex flow while ensuring physical consistency and high computational efficiency. The ESRGAN model has a strong information mapping capability, capturing fluctuating features from local flow fields of varying geometries and sizes. The model effectiveness in reconstructing the whole domain flow field is validated by comparing instantaneous velocity fields, flow statistical properties, and probability density distributions. Using laminar bluff body flow from Direct Numerical Simulation (DNS) as a priori case, the model successfully reconstructs the complete flow field from three non-overlapping limited regions, with flow statistical properties perfectly matching the original data. Validation of the power spectrum density (PSD) for the reconstruction results also proves that the model could conform to the temporal behavior of the real complete flow field. Additionally, tests using DNS turbulent channel flow with a friction Reynolds number ($Re_\tau = 180$) demonstrate the model ability to reconstruct turbulent fields, though the quality of results depends on the number of flow features in the local regions. Finally, the model is applied to reconstruct turbulence flow fields from Particle Image Velocimetry (PIV) experimental measurements, using limited data from the near-wake region to reconstruct a larger field of view. The turbulence statistics closely match the experimental data, indicating that the model can serve as a reliable data-driven method to overcome PIV field-of-view limitations while saving computational costs.

\end{abstract}


\maketitle

\nopagebreak[4]

\section{\label{sec:level1}Introduction}

In various fields of fluid flow research, including energy and power engineering, aerospace engineering, and climate forecasting, fluid modeling, control, and reconstruction are commonly addressed. However, actual fluid dynamics problems are complex and variable, often necessitating the use of typical and simplified flow cases, such as flow around bluff bodies and boundary layer flows, for research purposes \cite{yousif2021high}. Despite this simplification, studies on fluid visualization, simulation analysis, and characteristic statistics require a substantial amount of effective and accurate flow data. Therefore, obtaining a large amount of reliable and high-quality fluid flow data is crucial for improving the efficiency of various fluid-related systems.\par

Sources for generating fluid flow data encompass various methods. Firstly, experimental measurements involve using laboratory equipment and instruments to collect fluid data, such as wind tunnel experiments and water tank studies \cite{liang2020single}. Sensors like pressure sensors and flow meters provide real-time data on the fluid field, while technologies such as high-speed photography and laser measurement capture detailed motion characteristics \cite{xing2017high}. Particle Image Velocimetry (PIV) is a widely used technique in fluid mechanics, employing digital image processing to measure velocity fields \cite{liu2022experimental}. PIV experiments introduce tracer particles into the fluid, illuminated by laser sources, and track their positions in the flow field across successive frames to determine velocity vectors at different positions \cite{melling1997tracer}. This non-intrusive technique avoids sensor interference, is straightforward to operate, and delivers high spatial resolution data promptly \cite{liberzon2004xpiv}. However, PIV has limitations; accuracy in image detection and tracking hinges on tracer particle selection\cite{huang1993limitation}. For high-speed flows, PIV may lack temporal resolution, and especially its field of view remains restricted, necessitating stitching multiple fields together for comprehensive flow analysis, which is computationally intensive \cite{kahler2012resolution, scharnowski2020particle}. Meanwhile, advanced full-field measurement techniques like PIV/PTV (Particle Tracking Velocimetry) and whole-field Reynolds stress measurements demand sophisticated image processing and significant computational resources \cite{astarita2007analysis, rohacs2023past}.\par

Secondly, numerical simulations in fluid mechanics refer to Computational Fluid Dynamics (CFD) methods that simulate and predict fluid behavior. CFD discretizes partial differential equations into algebraic forms and solves them on computers to replicate actual fluid dynamics \cite{shang2004three}. Common methods include Reynolds-Averaged Navier-Stokes (RANS) \cite{alfonsi2009reynolds}, Large Eddy Simulation (LES) \cite{moeng1984large}, and Direct Numerical Simulation (DNS) \cite{moin1998direct}. Numerical simulations can model diverse fluid phenomena, including turbulence and multiphase flows, providing detailed information on flow variables like velocity, pressure, and temperature \cite{chirigati2021accurate}. Unlike experiments, simulations offer cost-effective, time-efficient trials in a virtual environment, enabling extensive parameter adjustments and model testing \cite{hu2024super}. Nonetheless, simulation accuracy heavily relies on model assumptions and computational resources, particularly for high-fidelity simulations demanding large-scale computing clusters or supercomputers \cite{fan2023high}.\par 

Other methods include field observations \cite{delannay2017granular}, data mining and analysis \cite{bellazzi2011data}, collaborative research and data sharing, and sensor networks and remote monitoring \cite{kumar2020reliable}.\par 

Commonly, fluid mechanics researchers use PIV experiments and CFD simulations to obtain accurate data, but addressing issues such as limited field of view and excessive computational resources remains challenging. Recent advancements in deep learning algorithms have shown considerable progress, being widely applied in fields such as medicine \cite{yu2023popular}, natural language processing \cite{otter2020survey}, image processing and computer vision \cite{wang2021person}, intelligent driving and robotics \cite{haydari2020deep}. With its powerful feature learning capabilities, deep learning model can also handle complex nonlinear, high-dimensional, and big data characteristic problems in fluid mechanics. Various deep learning-based methods have been applied to turbulence, such as turbulence modeling \cite{ling2016reynolds}, flow prediction \cite{lee2019data}, and flow control \cite{park2020machine}. More researchers are inclined to use deep learning methods to estimate high-resolution flow fields around physical models \cite{ling2016reynolds}. This task is similar to high-resolution reconstruction in computer vision. Reconstructing high-resolution velocity fields from low-resolution counterparts can also be viewed as a velocity field estimation process \cite{yu2023deep}. With the introduction of various algorithms based on high-resolution image reconstruction, high-resolution turbulence reconstruction has become a promising research topic in the field of fluid mechanics.\par

This study aims to explore the feasibility of using deep learning algorithms to address the limited field of view in PIV experimental data. In other words, it focuses on using a deep learning model to reconstruct complete flow field data from data based on limited regions. Since Rabault et al. (2017) proposed using shallow convolutional neural networks(CNNs) for PIV estimation \cite{rabault2017performing}, deep learning-based fluid motion estimation methods have continuously emerged \cite{cai2019dense}. Guemes et al. (2019) proposed a CNN model combined with proper orthogonal decomposition (POD) to reconstruct large-scale motions from wall shear stress \cite{guemes2019sensing}. Guastani et al. (2020) demonstrated that a fully CNN-based model could also produce good velocity field reconstructions from wall shear stress \cite{guastoni2020prediction}. Previous studies have shown that neural networks (NNs) can infer small-scale structures and reconstruct turbulent fields from low-resolution spatiotemporal measurements \cite{liu2020deep, kim2021unsupervised}. This type of super-resolution reconstruction technology is primarily used to improve the quality of experimental and simulation data under limited resolution conditions. \par

Meanwhile, recent studies have shown that CNN-based generative adversarial networks (GANs) can reconstruct three-dimensional (3D) velocity fields from 2D observations \cite{yousif2023deep}. Notably, Buzzicotti et al. (2021) first used a GAN to reconstruct 2D damaged snapshots of a 3D rotating turbulence, even with large gaps and missing large and small-scale features \cite{buzzicotti2021reconstruction}. Kim et al. (2021) found that unsupervised deep learning could reconstruct high-resolution turbulence from unpaired training data \cite{kim2021unsupervised}. They used CycleGAN, a cyclic consistent generative adversarial network, to reconstruct high-resolution velocity fields based on low-resolution DNS and LES data. Comparatively, this model's reconstruction accuracy was better than bilinear interpolation and CNN-based models. Tianyi Li et al. (2023) used linear and nonlinear tools based on POD and GAN to reconstruct snapshots of rotational turbulence with spatial damage, confirming that the GAN approach was superior to the linear POD technique when comparing statistical multiscale characteristics \cite{li2023multi}.\par

Especially concerning the reconstruction work using CNN-based models applied to PIV experimental data, Cai et al. (2019) proposed a CNN-based model for estimating high-resolution velocity fields in PIV measurements, showing better performance compared to traditional cross-correlation algorithms \cite{cai2019dense}. Morimoto et al. (2020) used a similar CNN-based model to estimate velocity fields using PIV measurement data with missing regions, demonstrating the model ability to estimate missing regions in the velocity field accurately \cite{morimoto2021experimental}. Deng et al. (2019) applied a super-resolution GAN (SRGAN) and enhanced SRGAN (ESRGAN) to reconstruct high-resolution flow fields using PIV measurements of flow around a cylinder, showing that both models could accurately reconstruct mean and fluctuating flow fields, with ESRGAN performing better than SRGAN \cite{ledig2017photo, deng2019super, wang2019esrgan, yousif2021high}.\par

After comparing the strengths and weaknesses of common full-field measurement techniques and neural network models, this study adopts an enhanced super-resolution generative adversarial network (ESRGAN)-based model to reconstruct a complete flow field from local limited regions data. This model represents an advanced solution to mitigate the inherent field-of-view constraints found in PIV fluid experiment data. The difficulty in flow field reconstruction lies in accurately capturing and reconstructing large amounts of data under nonlinear, multi-scale, and complex boundary conditions while ensuring physical consistency and high computational efficiency. The model, with its strong information mapping capability \cite{wang2019esrgan}, can capture fluctuation characteristics from local regions of various shapes and sizes. The model incorporates a residual-in-residual dense block (RRDB) and a relativistic discriminator \cite{wang2019esrgan}, increasing network depth and enhancing training, thereby improving reconstruction performance and reducing computational complexity and memory usage. \par

To assess the reconstruction performance of the ESRGAN model, three cases have been used. The priori case used 2D laminar flow data around a square cylinder to test based on different limited regions, confirming that the model could reconstruct complete flow fields from three non-overlapping limited regions, with flow statistics perfectly matching the original data. Secondly, based on more intricate 2D turbulent channel flow data at $Re_\tau = 180$, further tested the model's ability to reconstruct turbulent fields, though the results depended on the amount of flow features contained in the limited regions. The final case applied the model to reconstruct turbulent flow fields measured by PIV, using restricted data from the near-wake region to reconstruct larger field-of-view flow fields, with turbulence statistics closely matching experimental data. The results indicate that this model can serve as a data-driven method to address PIV field-of-view limitations, offering reliability and cost savings.\par

The remainder of this paper is organized as follows. Section~\ref{sec:Methodology} explains the basic structure of the ESRGAN model used in the study. Section~\ref{sec:Data description and pre-processing} lists the basic information of the datasets used for model training and testing in three cases, sourced from DNS and PIV experiments. Section~\ref{sec:Results and discussion} presents a comparative analysis of the reconstructed complete flow fields based on different local regions in each case. Section~\ref{sec:Conclusion} concludes the study and presents the findings.\par

\section{Methodology}
\label{sec:Methodology}

GANs model applications in fluid dynamics extend beyond image enhancement \cite{ledig2017photo}, demonstrating excellence in tasks such as experimental data denoising, high-resolution restoration, and reduced-order modeling. The proposed ESRGAN model in this study can reconstruct the complete flow field based on input data from the limited regions. This approach falls under the category of supervised learning (SL), where the paired data, i.e., limited regions data and the entire domain data, are input randomly to the SL model. The model ultimately learns to reconstruct the complete flow field. Reconstructing turbulent fields is particularly challenging due to two factors. Firstly, turbulence exhibits numerous active degrees of freedom that increase with turbulence intensity, typically parameterized by the Reynolds number. Secondly, there is spatio-temporal sparsity, dependent on the area and geometric shapes of the missing regions. Considering the limited field of view in PIV experiments, the shapes of these locally missing regions should resemble camera snapshots. When testing the model, only independent, untrained data from limited regions is needed to obtain the reconstructed complete flow field. Applying this method to practical PIV experiments can extend the experimental field of view and reduce computational complexity and costs.\par

Since the inception of the first version of GANs \cite{goodfellow2014generative}, a multitude of variants have emerged. Nevertheless, unlike traditional multi-layer perceptrons (MLPs) or CNN-based architectures, GAN architectures remain fundamentally consistent. In GANs, two adversarial networks, namely the generator ($G$) and discriminator ($D$), compete against each other. Here, $G$ is responsible for generating fake images that resemble real images, while $D$ acts as a sensor to distinguish between fake and real images. During training, $G$ continually optimizes its forgery level based on feedback from $D$, ultimately generating fake images that are difficult for $D$ to distinguish from real ones. By the way, the discriminative ability for $D$ is also enhanced. The whole process approximates a minimum two-player confrontation game, and the value function $V(D,G)$ is defined as follows:\par

\begin{align}
\min_G \max_D V(D,G) &= \mathbb{E}_{x_r \sim P_{data}(x_r)} [\log D(x_r)] + \mathbb{E}_{\theta \sim P_\theta(\theta)} [1 - \log D(G(\theta))] \tag{1}
\end{align}

In the formula, $x_r$ represents the ground truth data of the complete flow field, i.e., the real image, and $P_{data}(x_r)$ is the distribution of real images. The operationEdenotes computing the average of all samples in a training minibatch, which is used to calculate the loss function or gradient updates in gradient descent algorithms. In the second term on the right-hand side of equation (1), $\theta$ is a random vector used as input to $G$, and $G(\theta)$ is the output of $G$. $D$ indicates the probability that an image is real rather than generated by the generator. During training, $G(\theta)$ is expected to generate an image similar to the real image, maximizing the value of $D(G(\theta))$ as much as possible. At the same time, $D(x_r)$ gradually increases toward 1, while $D(G(\theta))$ decreases toward 0. This means that during the whole training process, the training objective of $G$ is to minimize $V(D,G)$, while the training objective of $D$ is the opposite. If successful, $G$ should be able to generate artificial images that are indistinguishable by $D$.\par 

\begin{figure*}[ht]
    \centering
    \includegraphics[angle=0, trim=0 0 0 0, width=0.9\textwidth]{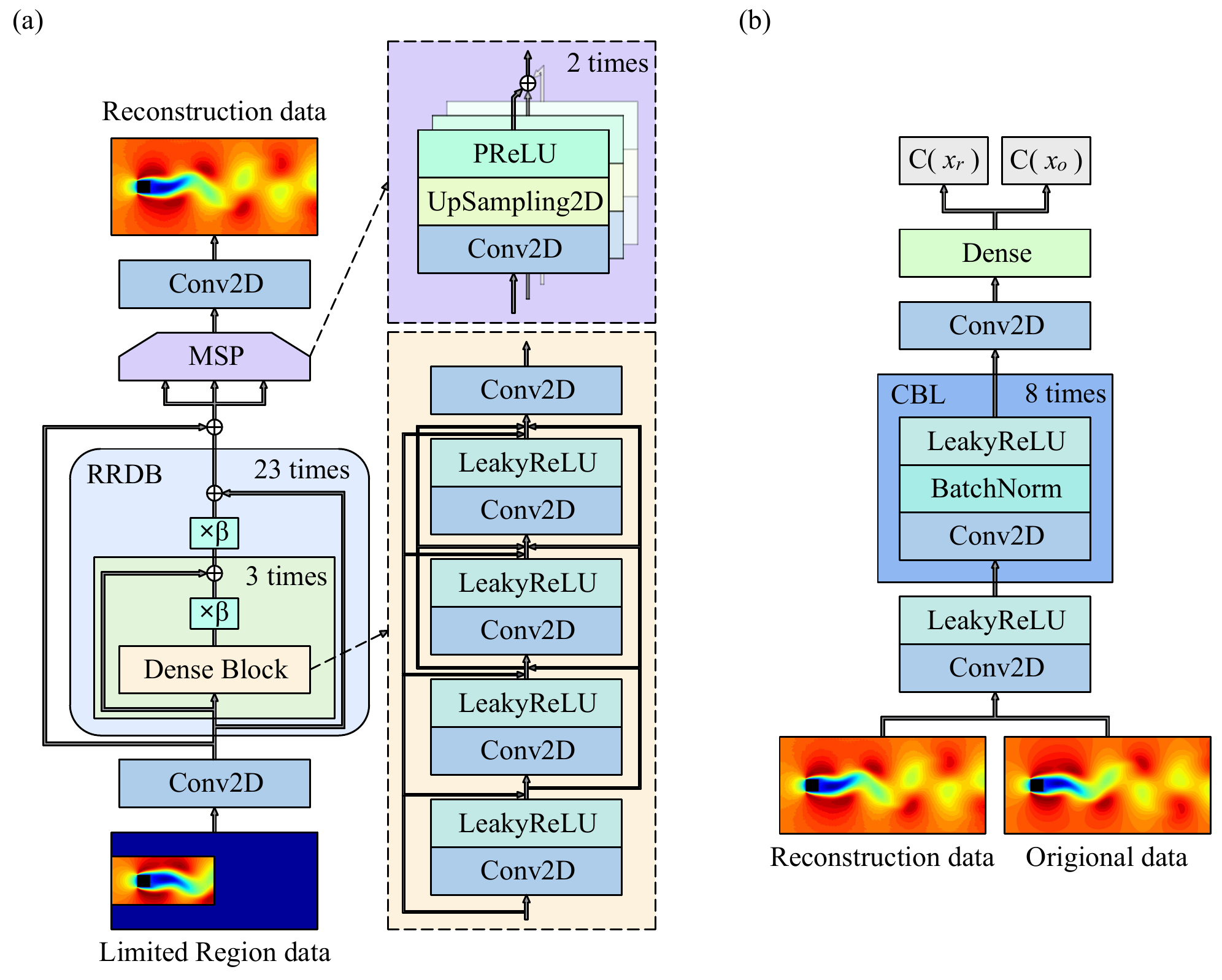}
    \caption[]{ESRGAN architecture: (a) the generator ($\beta$ is the residual scaling parameter = 0.2) and (b) the discriminator.}
    \label{fig:1}
\end{figure*}

The ESRGAN model used in this study is an Enhanced Super-Resolution GAN, capable of high-resolution data reconstruction \cite{wang2019esrgan}. Utilizing the high-fidelity deep learning network, it takes local restricted domain flow field data as part of the input and outputs reconstructed complete flow fields. Figure \ref{fig:1} illustrates the basic structure of the ESRGAN model. As shown in Figure \ref{fig:1} (a), $G$ comprises a deep convolutional neural network represented by residuals in residual dense blocks (RRDBs) and a multi-scale part (MSP). The RRDBs include more and deeper layers, functioning as a more efficient feature extractor. The input of local finite regions data enters the model, first transformed by a few convolutional layers and average pooling layers, and then passed through a series of consecutive RRDBs into the MSP\cite{yousif2021high}. The MSP consists of three parallel convolutional sub-models with different kernel sizes, as detailed in Table~\ref{tab:Table1}. Different kernel sizes imply different filter sizes, capturing various information. The data features extracted from RRDBs are reconstructed into complete flow field information through MSP. Subsequently, the outputs of the three branch sub-models are summed, followed by several convolutional layers, ultimately generating a complete flow field image ($x_f$).\par 

\begin{table}[ht]
    \centering
    \caption{MSP architecture.}
    \label{tab:Table1}
    \begin{tabular}{ccc}
        \hline
        First branch & Second branch & Third branch \\
        \hline
        Conv. (3, 3) & Conv. (5, 5) & Conv. (7, 7) \\
        UpSampling (2, 2) & UpSampling (2, 2) & UpSampling (2, 2) \\
        Conv. (3, 3) & Conv. (5, 5) & Conv. (7, 7) \\
        LeakyReLU & LeakyReLU & LeakyReLU \\
        UpSampling (2, 2) & UpSampling (2, 2) & UpSampling (2, 2) \\
        Conv. (3, 3) & Conv. (5, 5) & Conv. (7, 7) \\
        LeakyReLU & LeakyReLU & LeakyReLU \\
        UpSampling (2, 2) & UpSampling (2, 2) & UpSampling (2, 2) \\
        Conv. (3, 3) & Conv. (5, 5) & Conv. (7, 7) \\
        LeakyReLU & LeakyReLU & LeakyReLU \\
        \multicolumn{3}{c}{Add (first branch, second branch, third branch)} \\
        \hline
    \end{tabular}
\end{table}

Additionally, the structure of $D$ is illustrated in Figure \ref{fig:1} (b). Artificial fake images are fed into $D$ along with real images, and then passed through a series of CBL layers consisting of Convolution, Batch Normalization, and Leaky Rectified Linear Unit (LReLU). Finally, the data is crossed onto a convolutional layer\cite{yousif2021high}. The non-transformed discriminator outputs the real flow field image probability $C(x_r)$ and fake flow field image probability $C(x_f)$.\par 

In the training process, the generator takes as input local restricted domain data and artificially outputs fake whole domain flow field data. The discriminator receives real complete flow field data and fake complete flow field data, and computes the Relativistic average Discriminator (RaD) value using a non-transformed discriminator, denoted as $D_{Ra}$ \cite{wang2019esrgan, yousif2021high}, with the formula as follows: \par

\begin{equation}
D_{\text{Ra}}(x_r, x_f) = \sigma(C(x_r) - \mathbb{E}_{x_f}[C(x_f)])\tag{2}
\label{eq:2}
\end{equation}

\begin{equation}
D_{\text{Ra}}(x_f, x_r) = \sigma(C(x_f) - \mathbb{E}_{x_r}[C(x_r)])\tag{3}
\label{eq:3}
\end{equation}

In equations \ref{eq:2} and \ref{eq:3}, $\sigma$ represents the sigmoid function. $D_{Ra}$  measures the probability that the output of $D$ is more realistic when using real data compared to using fake data.\par

The discriminator loss function is defined as follows:\par

\begin{equation}
F_D = -\mathbb{E}_{x_r} [\log(D_{Ra}(x_r, x_f))] - \mathbb{E}_{x_f} [\log(1 - D_{Ra}(x_f, x_r))]\tag{4}
\label{eq:4}
\end{equation}

Conversely, the generator adversarial loss can be symmetrically formulated as:\par

\begin{equation}
L_G^{Ra} = -\mathbb{E}_{x_r} [\log(1 - D_{Ra}(x_r, x_f))] - \mathbb{E}_{x_f} [\log(D_{Ra}(x_f, x_r))]\tag{5}
\label{eq:5}
\end{equation}

In addition to the adversarial loss, two additional loss terms are used to form the generator composite loss function, as shown in the formula below:\par

\begin{equation}
F_G = L_G^{Ra} + \eta L_{pixel} + L_{gradient}\tag{6}
\label{eq:6}
\end{equation}

Here, $L_{pixel}$  is the mean squared error computed based on the pixel difference between the generated fake flow field data and real complete flow field data. $L_{gradient}$  represents the difference between the gradient values of the generated data and the gradient values of the real data, or between the gradient values of the fake data and a smoothed version (zero gradient). Here, $\eta$ is a coefficient used to balance the relative importance of different loss terms. Based on practical model training and tuning, $\eta$ is set to 2000. The training data is divided into batches, with each batch size set to 16. Meanwhile, the ESRGAN model uses the adaptive moment estimation (Adam) gradient descent optimization algorithm to update the weights \cite{yousif2021high}.\par

\begin{figure*}[ht]
    \centering
    \includegraphics[angle=0, trim=0 0 0 0, width=1.0\textwidth]{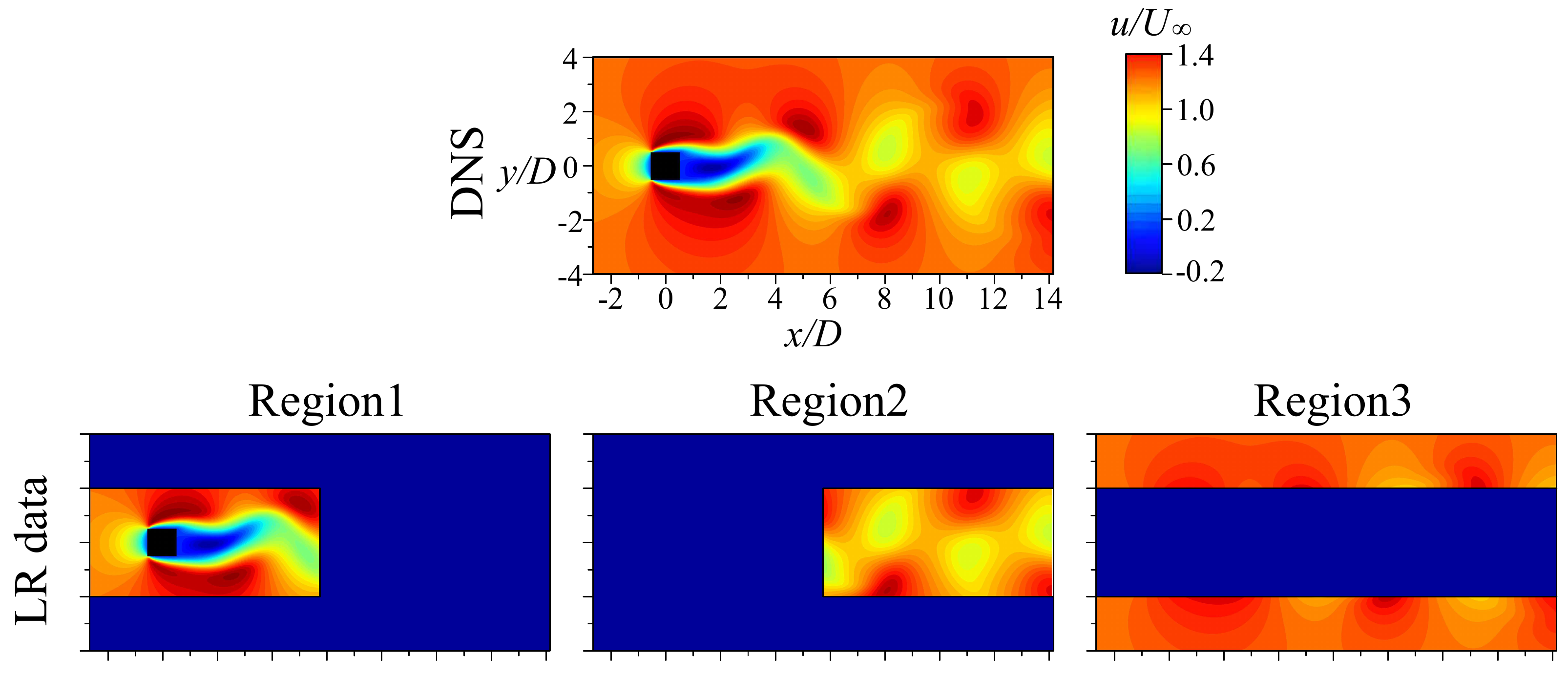}
    \caption[]{Schematic of dividing 2D laminar flow around a square cylinder into three regions. "LR" represents the limited regions data.}
    \label{fig:2}
\end{figure*}

\begin{figure*}[ht]
    \centering
    \includegraphics[angle=0, trim=0 0 0 0, width=1.0\textwidth]{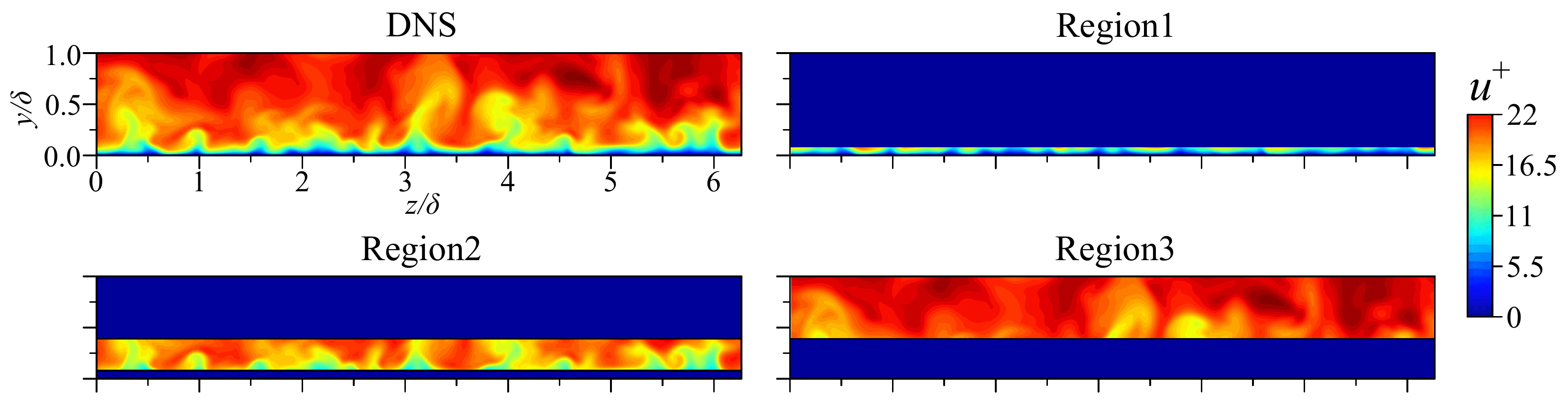}
    \caption[]{Schematic of dividing turbulent channel flow into three limited regions.}
    \label{fig:3}
\end{figure*}

\begin{figure*}[ht]
    \centering
    \includegraphics[angle=0, trim=0 0 0 0, width=0.9\textwidth]{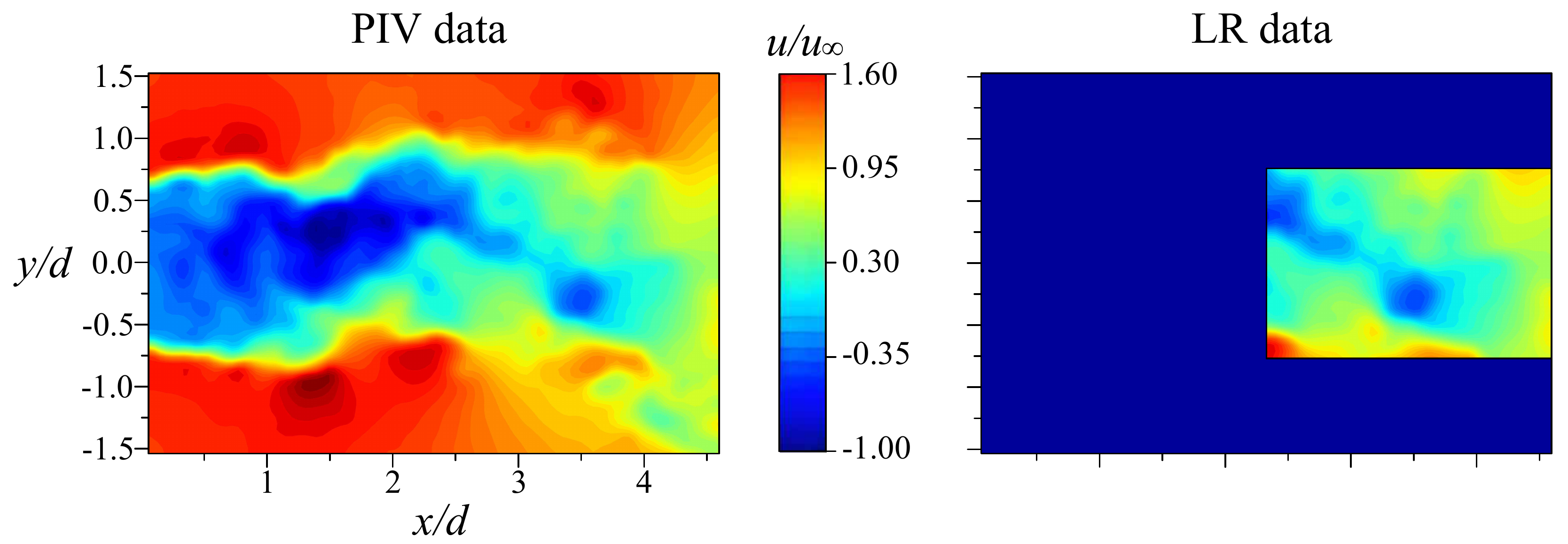}
    \caption[]{Schematic of dividing flow field around a circular cylinder into a limited region. "LR" represents the limited region data.}
    \label{fig:4}
\end{figure*}

\begin{figure*}[ht]
    \centering
    \includegraphics[angle=0, trim=0 0 0 0, width=0.9\textwidth]{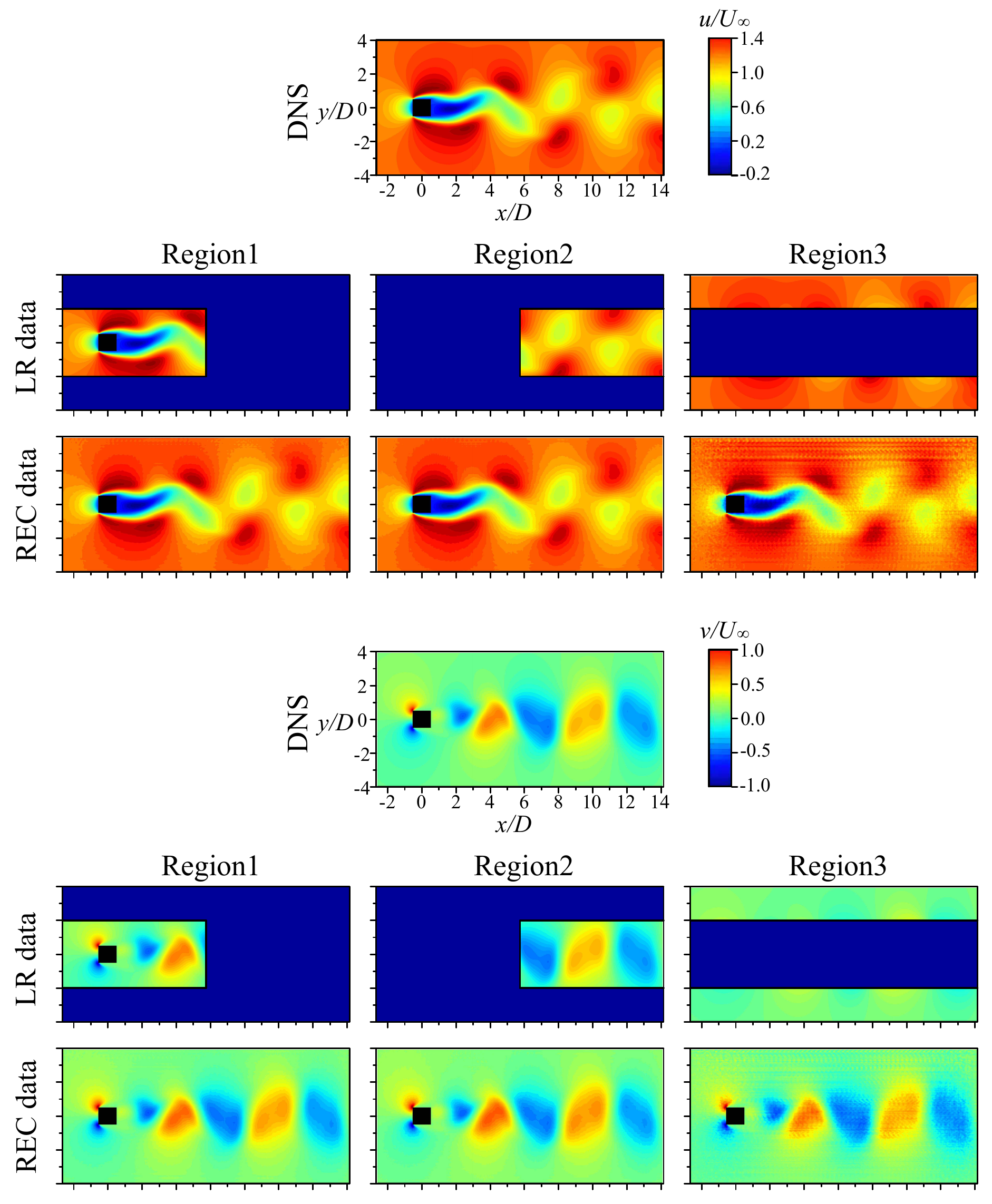}
    \caption[]{Reconstructed instantaneous velocity fields from three limited regions data. "LR" represents the limited regions data, "REC" represents the reconstruction data.}
    \label{fig:5}
\end{figure*}

\begin{figure*}[ht]
    \centering
    \includegraphics[angle=0, trim=0 0 0 0, width=0.9\textwidth]{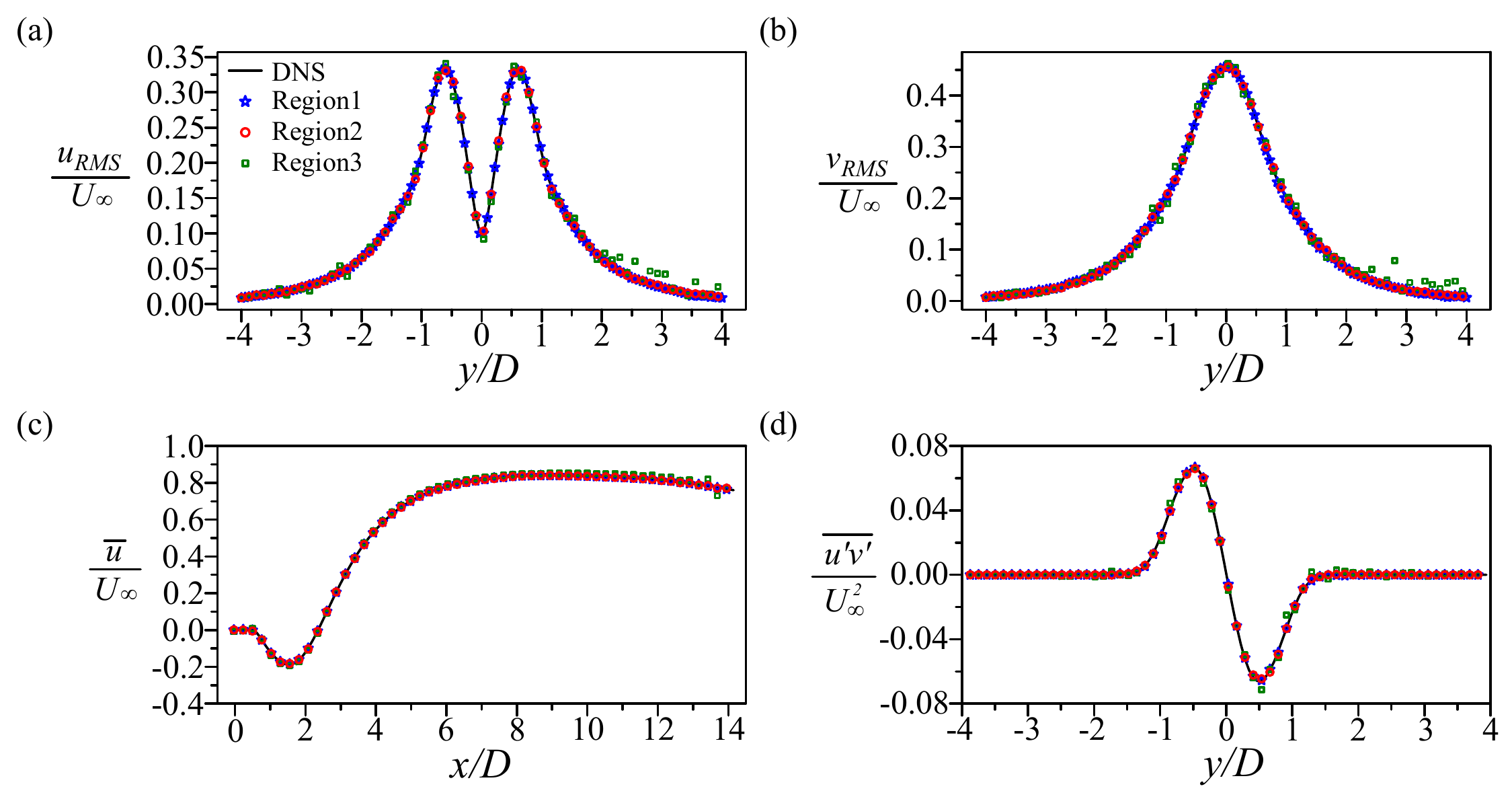}
    \caption[]{Profiles on the RMS of the (a)(b) velocities, (c) mean streamwise velocity, (d) Reynolds shear stress based on three limited regions reconstruction.}
    \label{fig:6}
\end{figure*}

\begin{figure*}[hpt]
    \centering
    \includegraphics[angle=0, trim=0 0 0 0, width=0.9\textwidth]{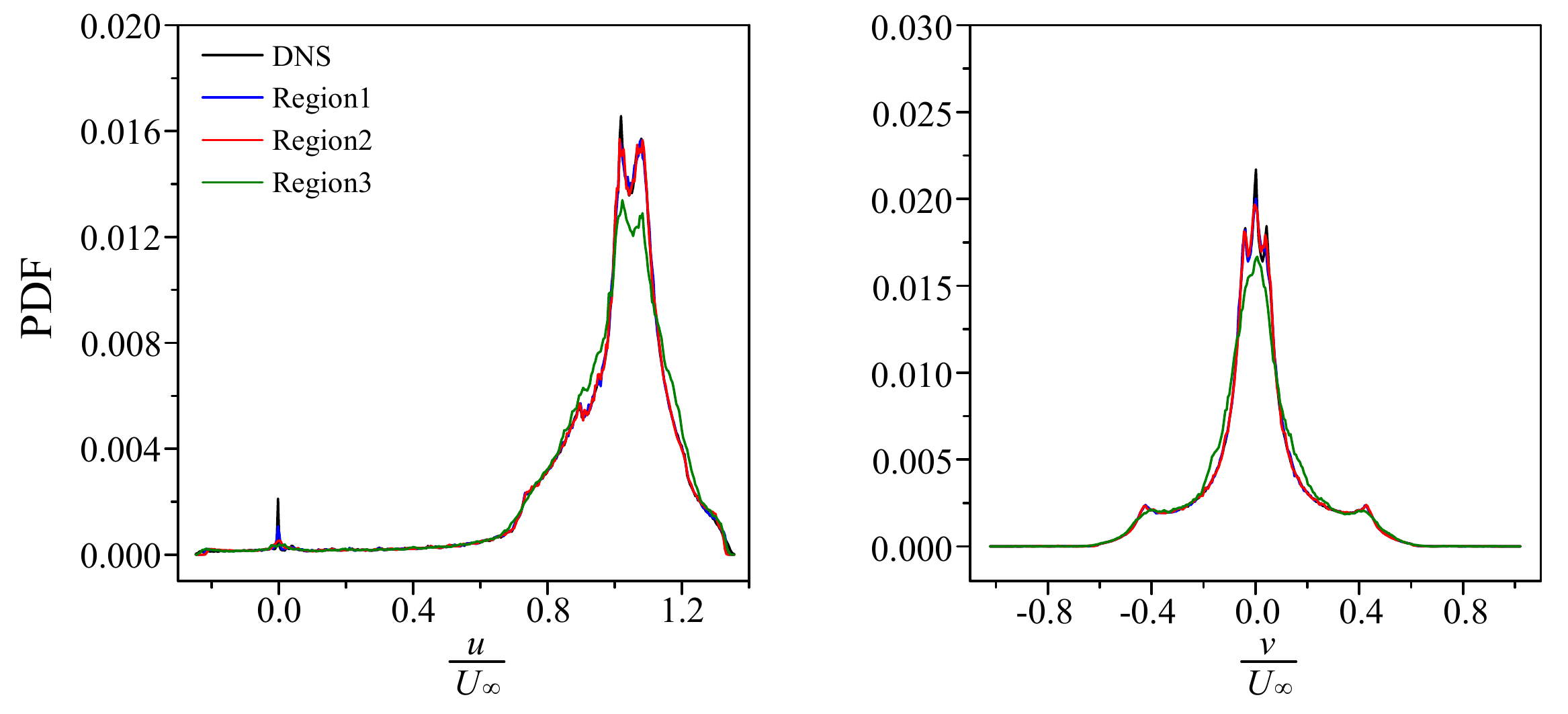}
    \caption[]{Probability density functions of the velocity components. The colored lines represent the data generated based on the reconstruction of the three limited regions.}
    \label{fig:7}
\end{figure*}

\begin{figure*}[hpt]
    \centering
    \includegraphics[angle=0, trim=0 0 0 0, width=0.9\textwidth]{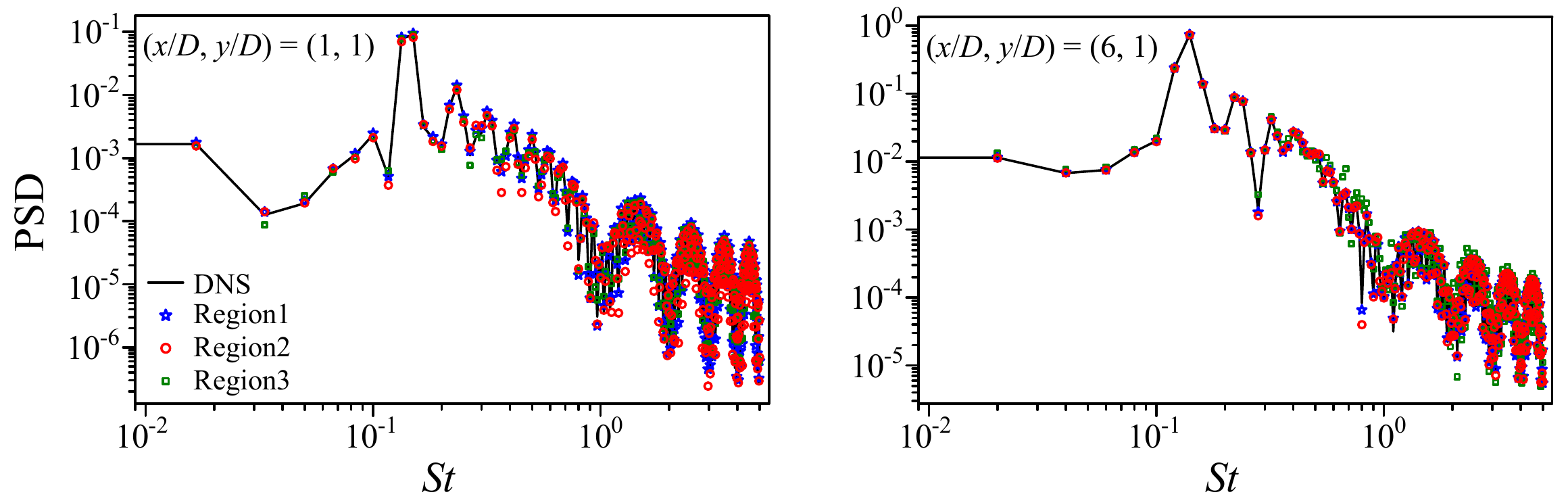}
    \caption[]{Power spectrum density of the reconstructed streamwise velocity fluctuations at two different streamwise locations.}
    \label{fig:8}
\end{figure*}

\section{Data description and pre-processing}
\label{sec:Data description and pre-processing}

This section introduces how the data for the three cases used in the study are generated and obtained through DNS and PIV experiment. It also explains the relevant data information and preprocessing steps for each case, including partitioning the complete flow data into several different sub-regions corresponding to whole domain flow and restricted domain flow data.\par

\subsection{Laminar flow around a square cylinder obtained by DNS}

The first case used to test the model involves a 2D laminar flow dataset around a square cylinder generated by DNS. The Reynolds number ${Re}_D = \frac{U_\infty D}{\nu} = 100$ is calculated based on the free-stream velocity $U_\infty$, cylinder width $D$, and fluid kinematic viscosity $\nu$. Using the open-source CFD software OpenFOAM-5.0x, a specific version, the detailed flow behavior around the square cylinder is simulated using the finite volume method. The complete simulation domain is characterized by coordinate axes representing streamwise ($x$), wall-normal ($y$), and spanwise ($z$) directions, with corresponding velocity components $u$, $v$, and $w$. The dimensions of the structural domain through which the laminar flow around the square cylinder passes are set as ($x_D, y_D$) = (20$D$, 15$D$), with a corresponding grid size of (381 × 221). The simulation employs two governing equations: the momentum equation (Equation \ref{eq:7}) for incompressible viscous fluid and the continuity equation (Equation \ref{eq:8}) representing fluid incompressibility. The equations are as follows:\par 

\begin{equation}
\frac{\partial \mathbf{u}}{\partial t} + \mathbf{u} \cdot \nabla \mathbf{u} = -\frac{1}{\rho} \nabla p + \nu \nabla^2 \mathbf{u} \tag{7}
\label{eq:7}
\end{equation}

\begin{equation}
\nabla \cdot \mathbf{u} = 0 \tag{8}
\label{eq:8}
\end{equation}

Here, $\mathbf{u}$ represents the fluid velocity, including components in the streamwise ($u$), wall-normal ($v$), and spanwise ($w$) directions. $t$, $\rho$, $p$, and $\nu$ represent time, density, pressure, and fluid kinematic viscosity, respectively. Stretching grid technology is used to locally refine the mesh around the square cylinder. Uniform inlet velocity and pressure outlet boundary conditions are set at the entrance and exit of the channel, respectively. The surface of the cylinder is subject to a no-slip wall boundary condition, with symmetrical boundaries on either side. The simulation time step is set to $\Delta t = 10^{-2}$, matching the dimensionless time step $\Delta t^* = \Delta t \frac{U_\infty}{D}$. Statistical data obtained from the simulation is validated against DNS results obtained by Anzai et al \cite{anzai2017numerical}. Based on this, the data used in this study is obtained by cropping the original simulation data around the edges and interpolating it onto a uniform grid. The new structural domain size is ($x_D, y_D$) = (17$D$, 8$D$) with a corresponding uniform grid size of (256 × 128). The time interval between flow snapshots is 0.1, which is 10 times the simulation time step. A total of 7000 complete flow field snapshots under laminar conditions were collected.\par   

To obtain different restricted domain data, the original complete flow snapshots are segmented into three local regions as shown in Figure \ref{fig:2}, "Region 1", "Region 2", and "Region 3". Considering the correspondence of fluid flow behavior on symmetrical surfaces, the upper and lower boundary area data are combined into "Region 3", equivalent to occupying 2 × (256 × 32) grid points. Additionally, to fairly balance the importance of characteristic information for area data around the square cylinder and the bluff body wake flow area, "Region 1" and "Region 2" are set to have the same number of grid points, both (128 × 64). Thus, three equally sized local restricted domain datasets are derived based on the original complete flow field data. Each type of local region data consists of 6000 snapshots for training and 1000 snapshots for testing, with no overlap between the training and testing datasets.\par 

Notably, the division of the three local area shapes is arbitrarily chosen without specific physical significance. The 2D velocity field (including $u$ and $v$) is used during both training and reconstruction processes. Training data is normalized using a min-max normalization function to produce values between 0 and 1, followed by randomized input.\par

\subsection{Turbulent channel flow obtained by DNS}

The second case also utilizes the open-source CFD finite volume code OpenFOAM-5.0x for DNS computations. It involves simulated 3D turbulent channel flow data, with a friction Reynolds number  $Re_\tau = \frac{u_\tau \delta}{\nu} = 180$, where $u_\tau$ is the friction velocity, $\delta$ is half the channel height, and $\nu$ is the fluid kinematic viscosity. Detailed simulation parameters for turbulent channel flow are listed in Table \ref{tab:Table2}, where $L$ and $N$ represent domain dimensions and grid numbers, respectively. Using a Cartesian coordinate system, periodic boundary conditions are applied in the streamwise ($x$) and spanwise ($z$) directions, while the upper and lower walls of the channel are set with no-slip boundary conditions. Similarly, the governing equations include the momentum equation for incompressible viscous fluid. The simulated turbulent statistics are validated by comparison with results from Kim et al. \cite{kim1987turbulence} and Moser et al. \cite{moser1999direct}.\par 

\begin{table*}[ht]
    \centering
    \label{tab:Table2}
    \caption{Simulations parameters of the turbulent channel flow. The superscript "$ + $" indicates that the quantity is nondimensionalized by $ u_\tau $ and $ \nu $. $ \Delta y_w^+ $ and $ \Delta y_c^+ $ are the spacing near the wall and at the center of the channel. $ \Delta t^+ $ is the dimensionless time step of the simulation.}
    \begin{tabular}{cccccccccc}
        \hline
        & $Re_\tau$ & $L_x\times L_y \times L_z$ & $N_x\times N_y \times N_z$ & $\Delta x^+$ & $\Delta z^+$ & $\Delta y_w^+$ & $\Delta y_c^+$ & $\Delta t^+$ & \\
        \hline
        & $180$ & $4\pi\delta\times2\delta\times2\pi\delta$ & $256 \times 128 \times 256$ & $8.831$ & $4.415$ & $0.63$ & $4.68$ & $0.113$ & \\
        \hline
    \end{tabular}
\end{table*}

A total of 9000 snapshots of the complete velocity and pressure fields in the $y - z$ plane of turbulent channel flow at $Re_\tau = 180$ are extracted at fixed time intervals. Local mesh refinement is applied near the upper and lower boundaries of the channel using stretching grid technology, achieved by selecting distributed points in the field to increase point density near the walls. Considering the symmetric nature of channel flow and to reduce the computational cost, only half of the original velocity field is used in this case, specifically the lower half of the channel flow data. The cropped structural domain size is $L_y$ × $L_z$ = $\delta$ × $2\pi\delta$, corresponding to grid numbers $N_y$ × $N_z$ = 64 × 256.\par

Similarly, as shown in Figure \ref{fig:3}, the data is still divided into three local regions. According to the boundary layer structure, in this case, $0 \leq y^+ < 14$ is identified as "Region 1", $14 \leq y^+ < 70$ is identified as "Region 2", and the remaining is considered as "Region 3". There are 8000 snapshots for reconstruction training and 1000 snapshots for testing the model. Missing information in the data is assigned a value of 0 to match the grid size of the lower half of the channel flow field. This data processing technique ensures that all three types of local region data are suitable inputs for the model.\par

\subsection{Flow field around a circular cylinder obtained by PIV}

Based on the validation results from the previous two cases and the practical constraints of the experimental field of view, the third case focuses on extracting data from the wake region behind the cylinder for reconstruction. It utilizes measured data from PIV experiments around a circular cylinder to capture 2D turbulent flow fields. The instantaneous turbulent flow fields in the $x - y$  plane are measured, with a data point distribution size of $N_x$ × $N_y$ = 59 × 87. To further analyze and process this data, the original flow field data undergoes resampling and interpolation to convert it onto a uniform grid. Then the data is adjusted to a grid size of 64 × 96, improving the spatial resolution.\par

In the PIV experiments measuring turbulent flow fields around a circular cylinder, a high-speed camera (FASTCAM Mini UX 50) and a continuous laser with a wavelength of 532 nm are used to construct a complete PIV system \cite{yousif2024flow}. Additionally, the circular cylinder model is fabricated from an acrylic plate that is not completely transparent visually. Therefore, when the laser passes through the model, it creates partial shadow areas beneath the bluff body, which minimally impacts the required measurement data for this experiment. Snapshots are captured at intervals of $\Delta t = 10^{-4}$ s, resulting in a total of 2000 instantaneous flow field snapshots. Of these, 1600 are used for training, while the remaining snapshots are reserved for testing. The Reynolds number ${Re}_d = \frac{u_\infty d}{\nu} = 6000$ is calculated based on a free stream velocity $u_\infty$ = 3 m/s, cylinder width $d$ = 0.03 m, and fluid kinematic viscosity $\nu$ = 1.5 × $10^{-5}$. In this case extracts local flow velocity data similar to "Region 2" in Case 1 as model input, with the data distribution size being 32 × 48, as shown in Figure \ref{fig:4}.\par

\begin{figure*}[ht]
    \centering
    \includegraphics[angle=0, trim=0 0 0 0, width=1.0\textwidth]{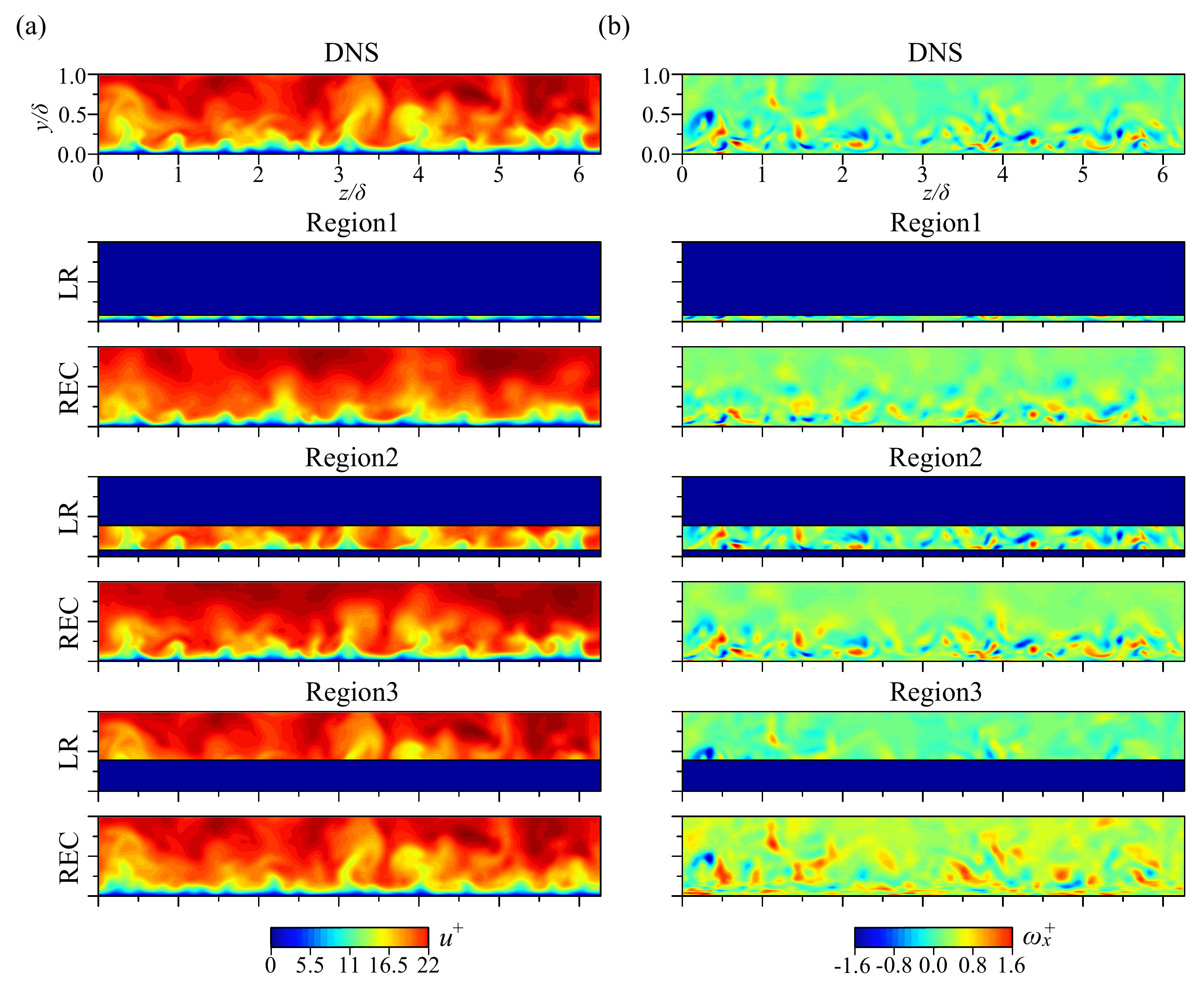}
    \caption[]{Reconstructed instantaneous streamwise (a) velocity and (b) vorticity fields from three limited regions data. The superscript "+" denotes the streamwise velocity and vorticity are nondimensionalized. "LR" represents the limited regions data, "REC" represents the reconstruction data.}
    \label{fig:9}
\end{figure*}

\begin{figure*}[ht]
    \centering
    \includegraphics[angle=0, trim=0 0 0 0, width=0.9\textwidth]{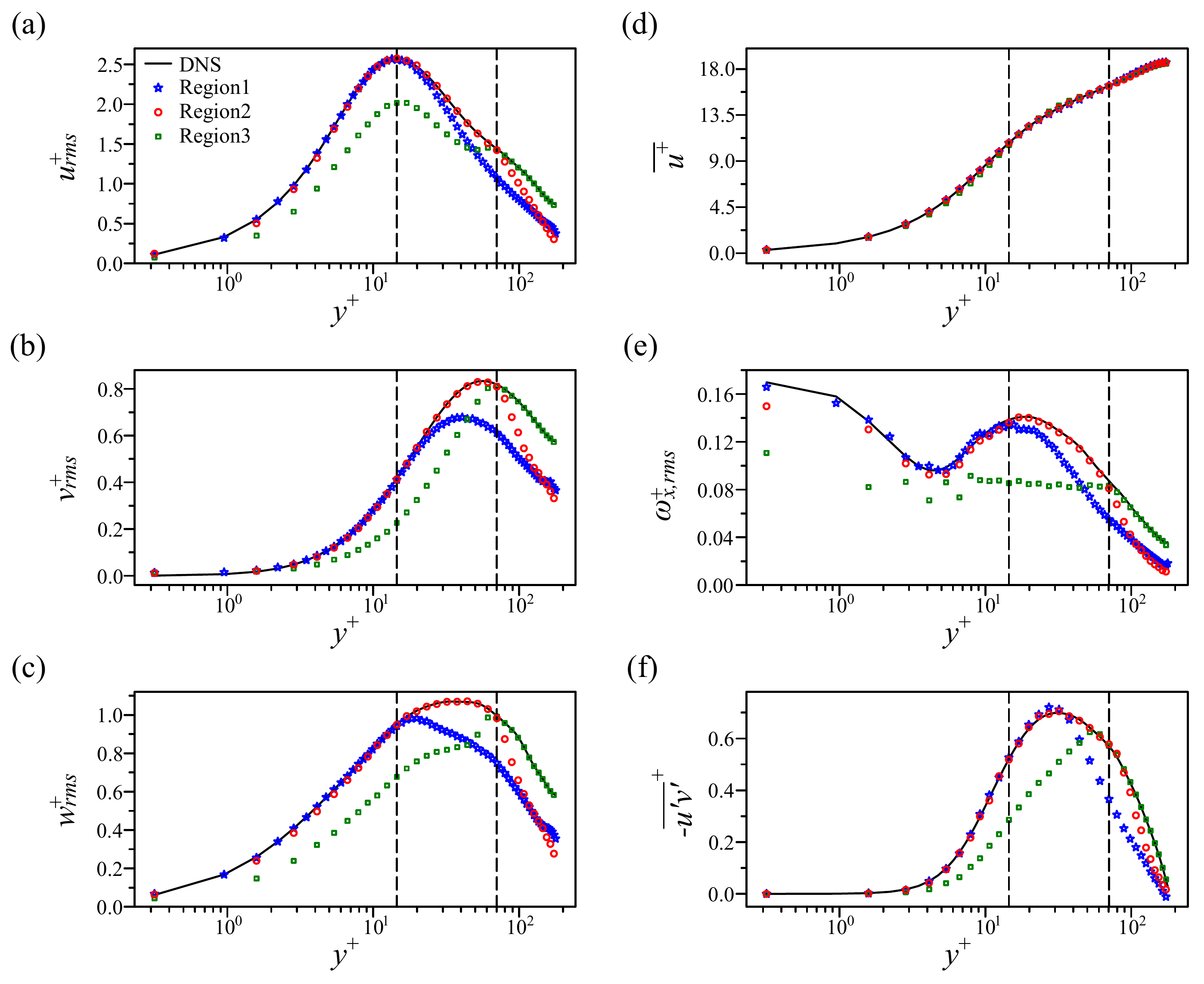}
    \caption[]{Profiles on the (a)(b)(c) RMS of the velocities \& (d) mean streamwise velocity \& (e) streamwise vorticity \& (f) Reynolds shear stress based on three limited regions reconstruction. The black dashed lines represent the boundarys of the regions division.}
    \label{fig:10}
\end{figure*}

\begin{figure*}[ht]
    \centering
    \includegraphics[angle=0, trim=5 5 0 0, width=0.9\textwidth]{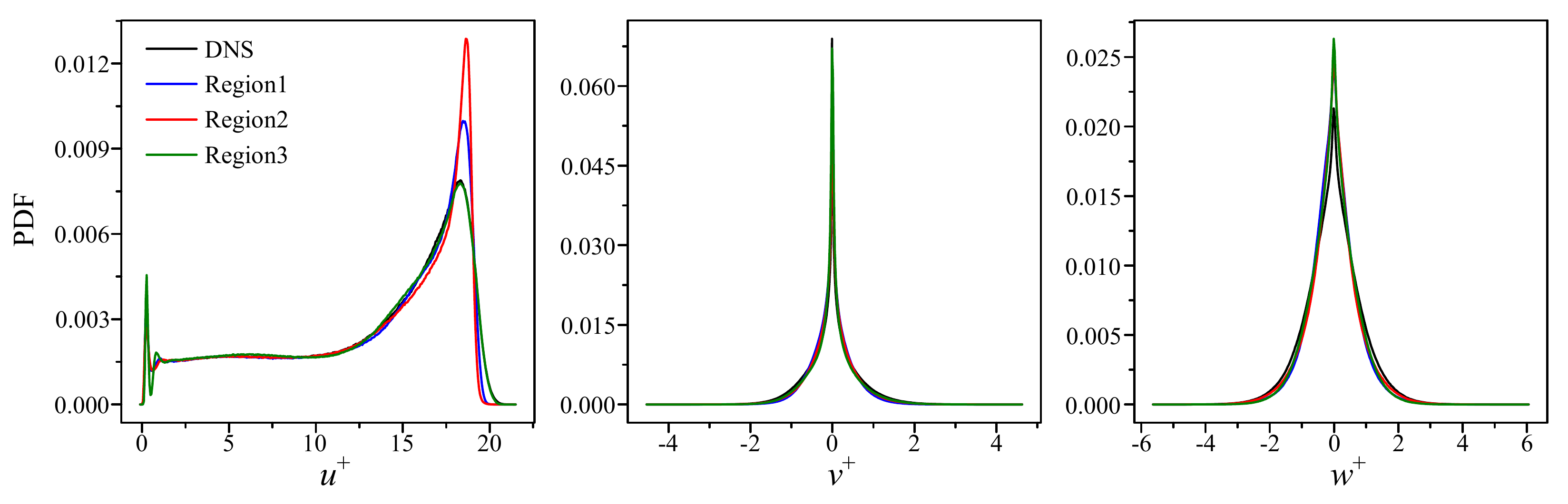}
    \caption[]{Probability density functions of the velocity components. The colored lines represent the data generated based on the reconstruction of the three limited regions.}
    \label{fig:11}
\end{figure*}

\begin{figure*}[ht]
    \centering
    \includegraphics[angle=0, trim=5 5 0 0, width=0.68\textwidth]{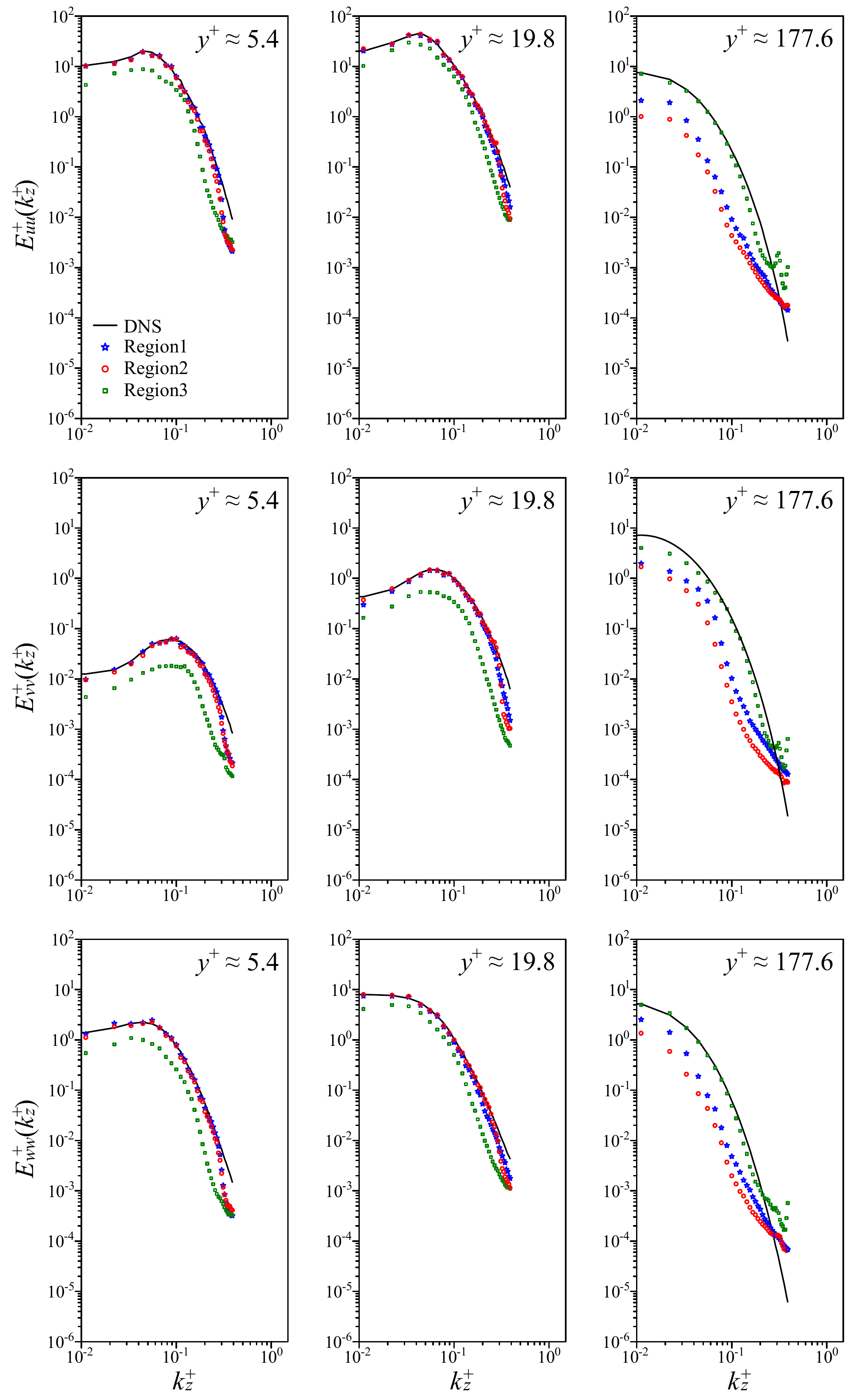}
    \caption[]{Spanwise energy spectra of the reconstructed velocity components for turbulent channel flow at $Re_\tau = 180$.}
    \label{fig:12}
\end{figure*}

\begin{figure}[ht]
    \centering
    \includegraphics[angle=0, trim=0 0 0 0, width=0.5\textwidth]{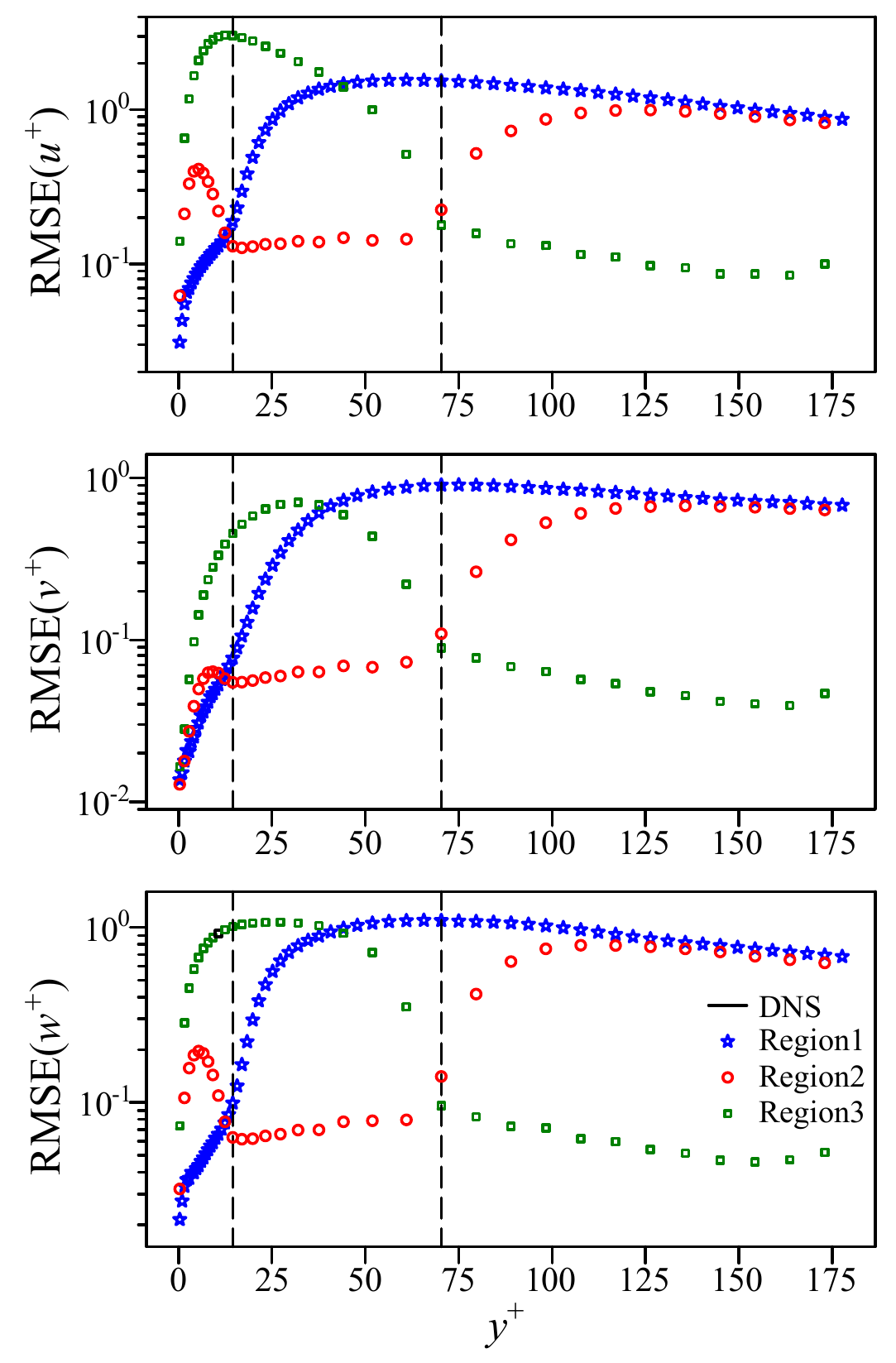}
    \caption[]{Profiles of RMSE for the reconstructed velocity components. The black dashed lines represent the boundarys of the regions division.}
    \label{fig:13}
\end{figure}

\begin{figure*}[ht]
    \centering
    \includegraphics[angle=0, trim=0 0 0 0, width=0.9\textwidth]{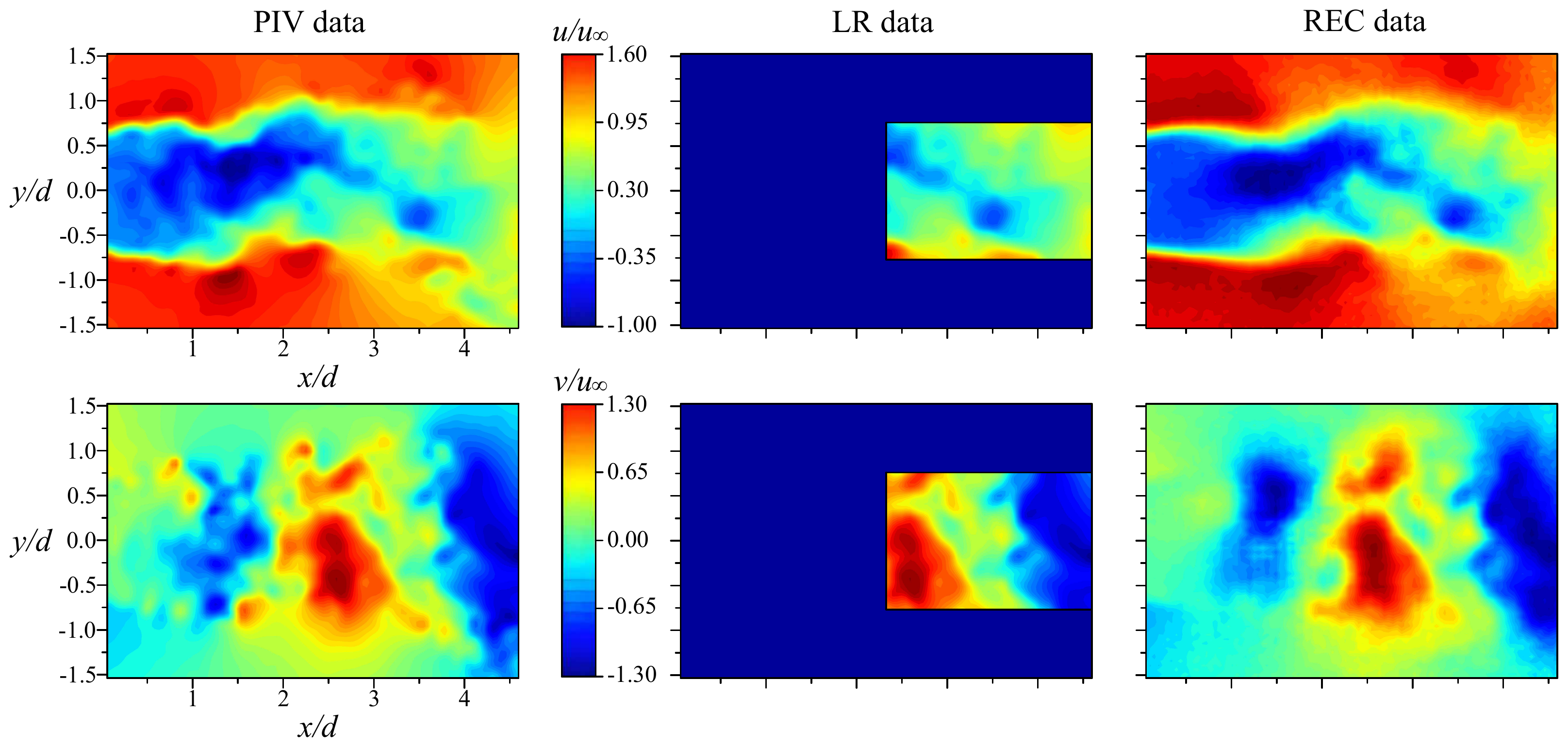}
    \caption[]{Reconstructed instantaneous velocity fields from limited region data obtained by PIV. "LR" represents the limited regions data, "REC" represents the reconstruction data.}
    \label{fig:14}
\end{figure*}

\begin{figure*}[ht]
    \centering
    \includegraphics[angle=0, trim=0 0 0 0, width=0.7\textwidth]{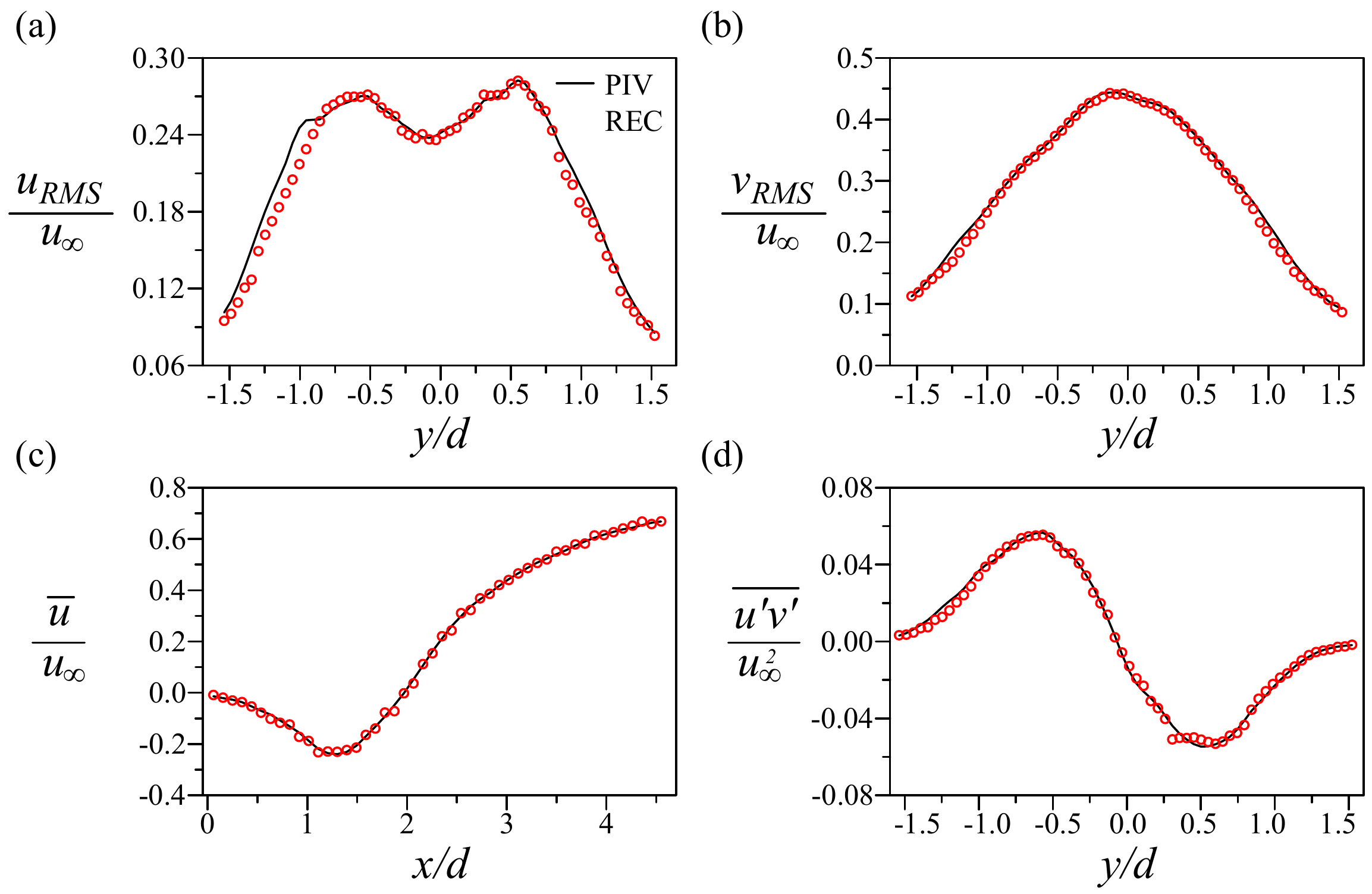}
    \caption[]{Profiles on the reconstructed (a)(b) RMS of the velocities \& (c) mean streamwise velocity \& (d) Reynolds shear stress from limited region data obtained by PIV.}
    \label{fig:15}
\end{figure*}

\begin{figure*}[ht]
    \centering
    \includegraphics[angle=0, trim=5 0 0 0, width=0.75\textwidth]{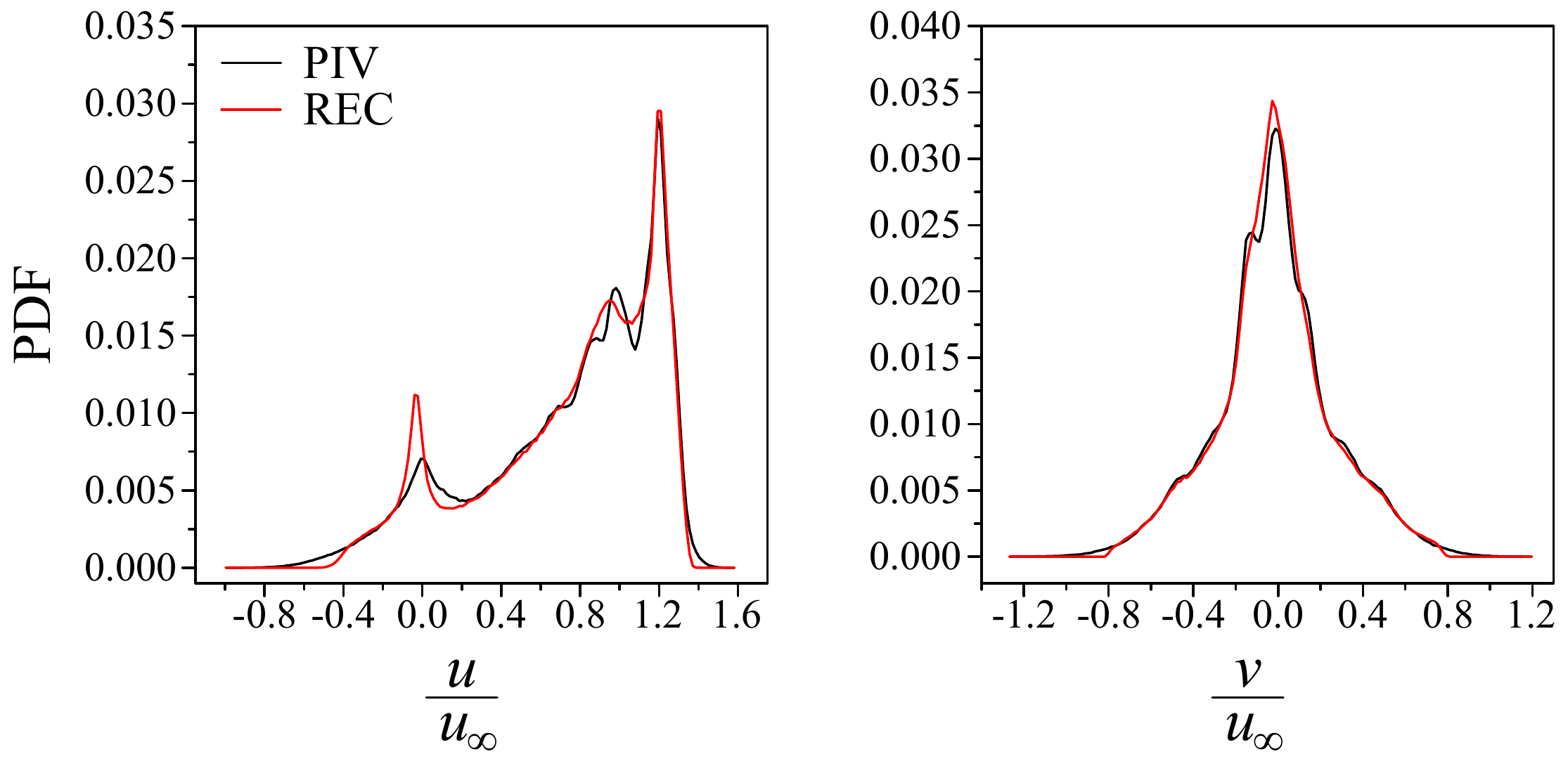}
    \caption[]{Probability density functions of the velocity components. The red lines represent the data reconstructed from limited region data obtained by PIV.}
    \label{fig:16}
\end{figure*}

\section{Results and discussion}
\label{sec:Results and discussion}

\subsection{Reconstruction of laminar flow around a square cylinder}

This section examines the ability of the ESRGAN model to reconstruct the complete laminar flow field based on different local regions. Here, testing data (not involved in the training process) is used to verify the model performance. The reconstructed instantaneous velocity fields based on three limited regions (LR) are shown in Figure \ref{fig:5}. The x-y coordinates in the figure are dimensionless by $D$ (cylinder width). As can be observed, the model can reconstruct the basic feature of the instantaneous flow field with high similarity to DNS data. However, the reconstruction performance on "Region 3" appears less effective compared to others, exhibiting some artifacts in the reconstructed complete flow field snapshots. Subsequent statistical analysis further confirms this phenomenon.\par

Figure \ref{fig:6} presents profiles of the root mean square (RMS) of the reconstructed velocity field, mean streamwise velocity, and Reynolds shear stress. Based on fixed cross-section observations, the RMS of both the streamwise and wall-normal velocities ($u_{RMS}$ and $v_{RMS}$) reconstructed from three limited regions are in good agreement with the DNS data. Only "Region 3" shows some minor deviations in the near-wake of the bluff body (i.e., at $x/D$ = 3 in the figure) when comparing the reconstructed velocity field with the original field in the wall-normal direction. This may be due to reduced zero-padding effects in the convolution process with increasing boundary layer thickness. Furthermore, the mean streamwise velocity ($\frac{\overline{u}}{U_\infty}$) and Reynolds shear stress ($\frac{\overline{u'v'}}{U_\infty^2}$) reconstructed from the three regions perfectly match the DNS data, confirming the model feasibility in recovering the complete velocity field.\par

Additionally, Figure \ref{fig:7} displays the probability density functions (PDF) of the reconstructed streamwise ($u$) and wall-normal ($v$) velocities field. Even the velocity PDF curve reconstructed from "Region 3" containing the least feature information, exhibits a relatively regular trend, although worse than other regions, especially around the cylinder and in the wake. The reconstruction results for "Region1" and "Region2" are almost equivalent. But the former leveraging data from the cylinder center (i.e., at $\frac{u}{U_\infty} = 0 $ and $\frac{v}{U_\infty} = 0 $ in the figure) performs slightly better here. Considering the zero-value assignment to the remaining regions during data processing, slight deviations from the DNS data in this case are understandable.\par

To further analyze the reconstruction results of the model based on data from these three limited regions, Figure \ref{fig:8} presents the power spectral density (PSD) of the streamwise velocity fluctuations in terms of the Strouhal number ($ St = \frac{fD}{U_\infty}$), where $f$ is the frequency. At these two different streamwise locations, the reconstruction results of "Region 1" and "Region 2" both align well with the DNS data. Even in "Region 3", where large-scale structures at low frequencies and small-scale activities at high frequencies are missing, the reconstruction results still broadly match the DNS data. This demonstrates that the reconstructed flow fields based on data from all three regions can conform to the temporal behavior of the real complete flow field.\par

There is clear evidence from the above demonstration results that the ESRGAN model can effectively reconstruct a complete laminar flow field while maintaining the same instantaneous and statistical results as the DNS data. Moreover, it is observed that the flow structures and feature information contained in "Region1" and "Region2" are significantly richer than those in "Region3".\par

\subsection{Reconstruction of turbulent channel flow }

This section presents the results of using the ESRGAN model to reconstruct turbulent channel flow data with a friction Reynolds number ($ Re_\tau = 180$). The model robustness is evaluated based on three non-overlapping limited regions. To match the grid size of the complete flow field, data outside the limited regions are assigned a value of 0. Considering the velocity distribution in the turbulent field, the model uses AveragePooling in the sampling layer for data input. \par

Figure \ref{fig:9} shows the reconstructed instantaneous streamwise velocity and vorticity fields obtained from three limited regions. The velocity field derived from "Region 1" visually deviates the most from the original DNS data. Conversely, "Region 3" with the largest proportion of original data, appears to have a better performance. "Region 2" exhibits the most prominent velocity fluctuation features, resulting in the best reconstruction. Overall, the primary characteristics of the complete flow field are accurately captured in the reconstruction results. However, the details of velocity fluctuations reconstructed based on the remaining areas without the original data are not clearly displayed. \par

The turbulent statistics of the reconstructed velocity fields are compared with the DNS turbulent channel flow in Figure \ref{fig:10}. Observations revealed that the statistical results of "Region 2" are the closest to the original DNS data, even though its data only accounts for five-sixteenths of the whole flow field grid. The lower boundary region reconstructed from this region aligns perfectly with the DNS data, only gradually deviating at the upper boundary. "Region 1" contains relatively few original data values, with some grid points having velocities close to 0. Additionally, during data preprocessing, the missing parts are assigned a value of 0, leading to small velocity fluctuations in the model input. In contrast, the average velocity value of "Region 3" is the highest, but it has fewer fluctuation characteristics, resulting in the assigned velocity field data almost directly jumping from 0 to the maximum value, hindering the model from capturing the correct characteristic information. Although the profile of the mean streamwise velocity shows that the model can reconstruct the complete flow field fairly well, the RMS of the streamwise vorticity and the Reynolds shear stress results significantly differ from the original DNS results. Considering that calculating these two characteristics involves errors in the wall-normal and spanwise velocities, such discrepancies are understandable.\par

Figure \ref{fig:11} presents PDFs of the velocity components, including streamwise velocity ($u$), wall-normal velocity ($v$), and spanwise velocity ($w$). The comparison indicates that the reconstructed wall-normal velocity generally matches well with the DNS data, while the performance on spanwise velocity is slightly inferior, with the poorest results observed in streamwise velocity. Interestingly, "Region 2", which shows the best reconstructed statistical characteristics, exhibits the largest deviation in high-speed streamwise velocity distribution. This is because the velocity values of "Region 2" are mostly concentrated at relatively high positions, showing greater fluctuation amplitude compared to the other two regions. However, the points with higher velocity values are mainly distributed in "Region 3", which includes the original DNS data. Meanwhile, "Region 1" primarily resides in the viscous sublayer, where 0 values dominate, exerting a weaker influence on high-speed streamwise velocities. \par

To further investigate the model's ability to reconstruct the complete velocity field with realistic behavior, the spanwise energy spectrum of each velocity component at different wall distances is studied based on data from three limited regions. As shown in Figure \ref{fig:12}, for flow with $Re_\tau = 180$, in all three regions, there is an appropriate shift at high wavenumbers, indicating errors in the reconstruction of small vortices. Notably, "Region 2" can reproduce a spectrum similar to DNS even when near-wall data is missing. Figure \ref{fig:13} shows the root mean square error (RMSE) of the velocity components reconstructed along the wall distance range. It can be seen that the reconstruction results of each local limited region perform best when the original DNS data is included. The RMSE of the reconstructed velocity components is mostly lower than 1, and all errors are within a reasonable range. As expected, the overall RMSE of "Region 2" is consistently lower than in other regions. This indicates that the buffer layer structure in "Region 2" is more stable, and the fluctuation characteristics are more significant, leading to higher accuracy. \par

As described above, based on the reconstruction results from these three limited regions, the ESRGAN model demonstrates a good capability to reconstruct turbulent flow fields. Additionally, the more velocity fluctuation features contained in the local regions, the clearer the details and the better the reconstruction results.\par

\subsection{Reconstruction of flow field obtained by PIV experiment}

Following the successful validation of the above two cases, the ESRGAN model is finally applied to address the limited field of view problem in PIV experiments. The section explores whether the model can reconstruct the complete 2D turbulent flow field based on the local region approximated at "Region 2". Due to experimental settings and equipment limitations, the initial PIV measurement data in this case cover a relatively small grid area, with only 2000 snapshots collected. This implies a very limited number of independent test data available for evaluating how the model performs on unknown data. \par

Figure \ref{fig:14} displays the PIV experimental and reconstructed instantaneous flow velocity fields, including streamwise velocity ($u$) and wall-normal velocity ($v$). Considering the shape and size of the field of view in PIV experiments, the local limited region approximates half of the complete flow field. Visually, the reconstructed velocity field resembles the results measured in the PIV experiment. This indicates that using partial near-wake data behind the cylinder, the model is able to reconstruct the morphology of the whole flow field. However, it exhibits some artifacts and lacks a few turbulent characteristics. This is due to the inherent noise errors in the experimental data and the limited amount of available training data.\par

The turbulent statistics of the reconstructed complete velocity field are shown in \ref{fig:15}. The RMS of the wall-normal velocity component ( $\frac{v_{RMS}}{u_\infty}$ ), mean streamwise velocity ( $\frac{\overline{u}}{u_\infty}$ ), and Reynolds shear stress ( $\frac{\overline{u'v'}}{u_\infty^2}$ )) all demonstrate that the ESRGAN model can restore the whole flow field quite well. However, due to insufficient data, the streamwise velocity RMS profile ( $\frac{u_{RMS}}{u_\infty}$ ) shows some unexpected deviations in the wall-normal direction. A small peak was observed in the profile measured by the PIV experiment. This is attributed to the limited number of snapshots used for testing, which resulted in inadequate feature learning by the model. Nevertheless, considering the overall comparison of the experimental flow field properties, the model reconstruction performance is still commendable. \par

Figure \ref{fig:16} presents the PDFs of the streamwise velocity ( $\frac{u}{u_\infty}$ ) and wall-normal velocity ( $\frac{v}{u_\infty}$ ) obtained from PIV measurement and reconstructed results. The comparison reveals that the reconstructed velocity field matches well with the velocity distribution from the PIV measurements. However, it is evident that the PDF curve for the wall-normal velocity matches the PIV data better than the PDF curve for the streamwise velocity. This means under the experimental conditions, fluid velocity changes in the wall-normal direction are more significant and measurable, whereas changes in the streamwise velocity might be more difficult to distinguish. Furthermore, due to slight velocity fluctuations near the zero value at the cylinder center, the reconstructed streamwise velocity PDF curve shows a small peak slightly higher than that of the PIV experimental data. \par

In summary, the ESRGAN model can indeed serve as a compensatory method for the limited field of view in PIV experimental measurements. This model can reconstruct complete flow field images based on data from local regions of various shapes and sizes. However, in practical applications, the training model does not have whole flow field data as input. The ESRGAN model still has capability to achieve complete flow field reconstruction, although the performance may be weakened. It can generalize to unknown region data, demonstrating a high interpolation potential.\par

\section{Conclusion}
\label{sec:Conclusion}

This study utilizes an Enhanced Super-Resolution Generative Adversarial Network (ESRGAN) model to reconstruct the whole domain flow field based on spatially restricted data. The model possesses powerful data information mapping capabilities, enabling it to capture fluctuation characteristics from local flow fields of various geometric shapes and sizes. During the model testing process, evaluations are primarily conducted from the perspectives of instantaneous velocity fields, flow statistical properties, and probability density functions.\par

To test the model reconstruction performance, three cases are used in the study. First, the model is tested using the 2D laminar flow around a square cylinder obtained from DNS. Complete instantaneous laminar flow fields were successfully reconstructed from three non-overlapping local regions, including "Region 3", which contained the fewest velocity fluctuation features. Comparison with DNS data showed that the reconstructed flow fields are consistent with the original ones, with negligible error. In the second step, the model robustness was further evaluated using DNS turbulent channel flow with a friction Reynolds number of $Re_\tau = 180$. The more velocity fluctuation features included in the local region, the clearer the details and better the reconstruction can be achieved. Despite the challenges posed by the high nonlinearity, multi-scale nature, complex dynamics of the turbulent field, and the influence of setting missing region values to zero, the model was still able to reconstruct the basic morphology and statistical characteristics of the whole turbulent field. Finally, the model is used to reconstruct a 2D turbulent field around a cylinder measured by PIV experiments. Using only a small amount of data from the wake region behind the cylinder center, the model was able to reconstruct the instantaneous velocity components of the complete field of view. In addition to visual analysis, statistical analysis of the velocity field also produced accurate results, confirming the model's reliable reconstruction capabilities.\par

In conclusion, the ESRGAN model used in this study proved to be a valuable supplement for expanding the view field in PIV experimental measurements. Its powerful reconstruction capabilities can be widely applied to meet visualization needs across various industries, including aerospace, medical imaging, and natural phenomenon prediction. In these fields, the precise data reconstruction and superior generalization capabilities of models are crucial.\par

\begin{acknowledgments}
This work was supported by 'Human Resources Program in Energy Technology' of the Korea Institute of Energy Technology Evaluation and Planning (KETEP), granted financial resource from the Ministry of Trade, Industry \& Energy, Republic of Korea (no. 20214000000140). This work was supported by the Korean Cancer Research Institute grant (2024).
\end{acknowledgments}

\section*{Data Availability}
The data that supports the findings of this study are available within this article.

\bibliography{2024ArXiv_MY_etal}

\end{document}